\documentclass[preprint,preprintnumbers,amssymb,prb,superscriptaddress]{revtex4}
\usepackage{graphicx,rotating}

\begin{document}

\title{Graphene: new bridge between condensed matter physics and quantum electrodynamics}
\author{M. I. Katsnelson}
\affiliation{Institute for Molecules and Materials, Radboud
University Nijmegen, 6525 ED Nijmegen, The Netherlands}
\author{K. S. Novoselov}
\affiliation{Manchester Centre for Mesoscience and Nanotechnology,
University of Manchester, Manchester M13 9PL, UK}

\begin{abstract}
Graphene is the first example of truly two-dimensional crystals -
it's just one layer of carbon atoms. It turns out to be a gapless
semiconductor with unique electronic properties resulting from the
fact that charge carriers in graphene demonstrate
charge-conjugation symmetry between electrons and holes and
possess an internal degree of freedom similar to ``chirality'' for
ultrarelativistic elementary particles. It provides unexpected
bridge between condensed matter physics and quantum
electrodynamics (QED). In particular, the relativistic
Zitterbewegung leads to the minimum conductivity of order of
conductance quantum $e^2/h$ in the limit of zero doping; the
concept of Klein paradox (tunneling of relativistic particles)
provides an essential insight into electron propagation through
potential barriers; vacuum polarization around charge impurities
is essential for understanding of high electron mobility in
graphene; index theorem explains anomalous quantum Hall effect.
\vspace{1cm}

\end{abstract}


 \maketitle

\section*{Introduction}
The variety of crystallographic forms of carbon places this
element into the focus of attention both in terms of basic
research as well as applications. The \textit{tree-dimensional}
crystallographic forms - graphite and diamond - are known from the
ancient times, and are widely used in industrial applications.
Recently discovered \textit{zero-dimensional} (fullerenes or cage
molecules~\cite{fulleren1,fulleren2,fulleren3}) and
\textit{one-dimensional} (carbon nanotubes~\cite{nanotube}) forms
are now extensively studied due to its' remarkable and, often,
unique mechanical and electronic properties. At the same time,
despite very intensive research in the area, no any
\textit{two-dimensional} form of carbon has been known until very
recently.

Ironically, this elusive two-dimensional form (dubbed graphene),
is, probably, the best theoretically studied carbon allotrope.
Graphene - planar, hexagonal arrangements of carbon atoms has been
the starting point in all calculations on graphite, carbon
nanotubes and fullerenes since late 40s~\cite{wallace}. However,
its' experimental discovery has been postponed till 2004 when a
technique called micromechanical cleavage has been employed to
obtain first graphene crystals~\cite{kostya0,kostya1}. The
observation of a peculiar spectrum of charge carriers and
anomalous quantum Hall effect (QHE) in graphene~\cite{kostya2,kim}
has initiated enormously growing interest to this field (for
review, see Refs.~\cite{reviewGK,reviewktsn}).

One of the most interesting aspects of the physics of graphene is
that it provides a novel example of a ``feedback'' of condensed
matter and material science on the fundamental physics. Of course,
this is not an unique case; archetypical examples are the concept of
spontaneously broken symmetry playing a crucial role in modern high
energy physics and quantum field theory~\cite{perkins} and the use
of M\"{o}ssbauer effect to check the general relativity
theory~\cite{pound}. At the same time, such relations are rather
rare and always turn out to be very fruitful. Actually, discovery of
graphene has opened new ways to study some basic quantum
relativistic phenomena which have always been considered as very
exotic. Probably the most clear example is the Klein
paradox~\cite{klein,dombey}, that is, a property of relativistic
quantum particles to penetrate with a probability of the order of
unity through very high and broad potential barriers. Previously it
was discussed only for experimentally unattainable (or very hard to
reach) situations such as particle-antiparticle pair creation at the
black hole evaporation, or vacuum breakdown at collisions of
super-heavy nuclei. At the same time, it appeared to be relevant for
graphene-based electronics~\cite{ktsn}. Here we discuss similarities
and differences between physics of charge carriers in graphene and
quantum electrodynamics (QED).

\section{Electronic structure of graphene}

\begin{figure}[t]
\begin{center}\leavevmode
\includegraphics[width=1\linewidth]{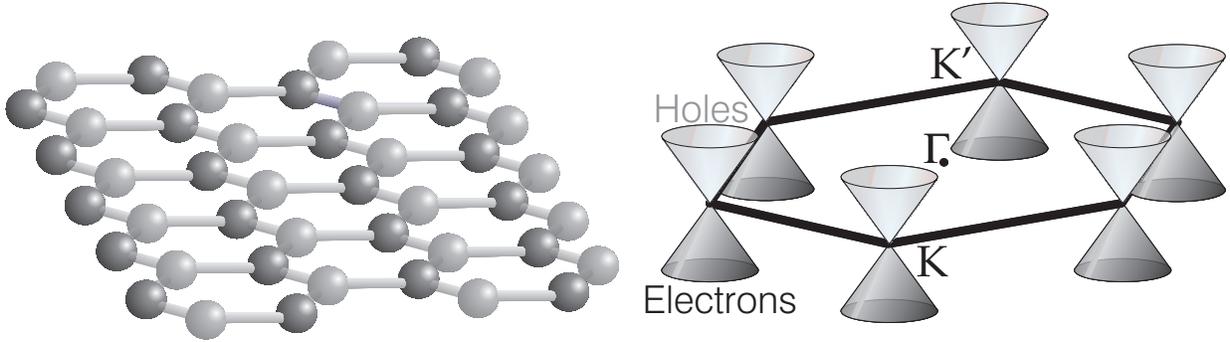}
\caption{Left: Crystallographic structure of graphene. Atoms from
different sublattices (A and B) are marked by different shades of
gray. Right: Band structure of graphene in the vicinity of the Fermi
level. Conductance band touches valence band at $K$ and $K'$ points.
} \label{GrapheneBand6}
\end{center}
\end{figure}

From the point of view of its electronic properties, graphene is a
two-dimensional zero-gap semiconductor with the energy spectrum
shown schematically in Fig.~\ref{GrapheneBand6} and its low-energy
quasiparticles formally described by the Dirac-like
Hamiltonian~\cite{slon,semenoff,haldane}
\begin{equation}
\widehat{H}_{0}=-i\hbar v_{F}\mathbf{\sigma }\nabla \label{zero}
\end{equation}
where $v_{F}$ $\approx10^{6}$ ms$^{-1}$ is the Fermi velocity, and
$\mathbf{\sigma=}\left( \sigma_{x},\sigma _{y}\right)$ are the
Pauli matrices. Neglecting many-body effects, this description is
accurate theoretically~\cite{slon,semenoff,haldane} and has also
been proven experimentally~\cite{kostya2,kim} by measuring the
energy-dependent cyclotron mass in graphene (which yields its
linear energy spectrum) and, most clearly, by the observation of a
relativistic analogue of the integer QHE which will be discussed
below.

The fact that charge carriers in graphene are described by the
Dirac-like equation (\ref{zero}) rather than the usual
Schr\"{o}dinger equation can be seen as a consequence of graphene's
crystal structure, which consists of two equivalent carbon
sublattices A and B~\cite{slon,semenoff,haldane} (see
Fig.~\ref{GrapheneBand6}). Quantum mechanical hopping between the
sublattices leads to the formation of two energy bands, and their
intersection near the edges of the Brillouin zone yields the conical
energy spectrum near the ``Dirac'' points $K$ and $K'$. As a result,
quasiparticles in graphene exhibit the linear dispersion relation $E
= \hbar k v_F$, as if they were massless relativistic particles,
with the role of the speed of light played by the Fermi velocity
$v_F\approx c/300$. Due to the linear spectrum, one can expect that
graphene's quasiparticles behave differently from those in
conventional metals and semiconductors where the energy spectrum can
be approximated by a parabolic (free-electron-like) dispersion
relation.

Although the linear spectrum is important, it is not the only
essential feature that underpins the description of quantum
transport in graphene by the Dirac equation. Above zero energy, the
current carrying states in graphene are, as usual, electron-like and
negatively charged. At negative energies, if the valence band is not
completely filled, its unoccupied electronic states behave as
positively charged quasiparticles (holes), which are often viewed as
a condensed-matter equivalent of positrons. Note however that
electrons and holes in condensed matter physics are normally
described by separate Schr\"{o}dinger equations, which are not in
any way connected (as a consequence of the Seitz sum rule~\cite{VK},
the equations should also involve different effective masses). In
contrast, electron and hole states in graphene are interconnected,
exhibiting properties analogous to the charge-conjugation symmetry
in QED~\cite{semenoff,haldane}. For the case of graphene, the latter
symmetry is a consequence of its crystal symmetry because graphene's
quasiparticles have to be described by two-component wavefunctions,
which is needed to define relative contributions of sublattices A
and B in the quasiparticles' make-up. The two-component description
for graphene is very similar to the one by spinor wavefunctions in
QED but the ``spin'' index for graphene indicates sublattices rather
than the real spin of electrons and is usually referred to as
pseudospin $\sigma$.

There are further analogies with QED. The conical spectrum of
graphene is the result of intersection of the energy bands
originating from sublattices A and B (see
Fig.~\ref{GrapheneBand6}) and, accordingly, an electron with
energy $E$ propagating in the positive direction originates from
the same branch of the electronic spectrum as the hole with energy
$-E$ propagating in the opposite direction. This yields that
electrons and holes belonging to the same branch have pseudospin
$\sigma$ pointing in the same direction, which is parallel to the
momentum for electrons and antiparallel for holes. This allows one
to introduce chirality~\cite{haldane}, that is formally a
projection of pseudospin on the direction of motion, which is
positive and negative for electrons and holes, respectively. The
term ``chirality'' is often used to refer to the additional
built-in symmetry between electron and hole parts of graphene's
spectrum and is analogous (although not completely
identical~\cite{semenoff,npb}) to the chirality in
three-dimensional QED.

An alternative view on the origin of the chirality in graphene is
based on the concept of ``Berry phase''~\cite{berry}. Since the
electron wave function is a two-component spinor, it has to change
sign when the electron moves along the close contour. Thus the
wave function gains an additional phase $\phi = \pi$.

The analogy with the field theory takes a very interesting twist,
should we take into account that a sheet of graphene must always
be corrugated. The fact that graphene crystals always exhibit some
finite local curvature can be considered as a consequence of the
Mermin-Wagner theorem, and has been confirmed experimentally both
for graphene samples resting on a substrate~\cite{morozov}, as
well as for free-hanging graphene films~\cite{jannik}.

It is well known that harmonic approximation in the
two-dimensional case doesn't produce a solution with long-range
order ~\cite{peierls1,peierls2,landau,LL}. One can see this as
bending instabilities, due to soft long-wavelength phonons, lead
to membrane crumpling~\cite{nelson}. Anharmonic coupling between
bending and stretching modes changes the situation drastically and
prevents the crumpling~\cite{nelson,peliti,radz}. However, the
membrane should be rippled in a sense that typical fluctuations in
the direction perpendicular to the surface $h\left( x,y\right)$
has a scale of order of $a\left( L/a\right)^{\zeta }\gg a$ where
$a$ is the lattice constant, $L$ is the size of the sample and
$\zeta $ is the roughness exponent. The latter can be estimated as
$\zeta \simeq 0.6$~\cite{nelson,radz}. For a typical sample size
$L\sim1\mu$m, the typical amplitude of the corrugation for
free-hanging membrane at room temperature was estimated as 0.5 nm,
with their characteristic size being about $5$nm ~\cite{jannik}.

These ripples lead to important consequences for the electronic
structure of graphene. The nearest-neighbor hopping integral
$\gamma$ turns out to be fluctuating due to its dependence on the
deformation tensor~\cite{nelson}
\begin{equation}
\overline{u}_{ij}=\frac{1}{2}\left( \frac{\partial u_{i}}{\partial x_{j}}+%
\frac{\partial u_{j}}{\partial x_{i}}+\frac{\partial u_{k}}{\partial x_{i}}%
\frac{\partial u_{k}}{\partial x_{j}}+\frac{\partial h}{\partial x_{i}}\frac{%
\partial h}{\partial x_{j}}\right)   \label{metric}
\end{equation}
where $x_{i}=\left( x,y\right) $ \ are coordinates in the plane
and $u_{i}$ are corresponding components of the displacement
vector:
\begin{equation}
\gamma =\gamma _{0}+\left( \frac{\partial \gamma }{\partial \overline{u}_{ij}%
}\right) _{0}\overline{u}_{ij}.  \label{depend}
\end{equation}
Taking into account this inhomogeneity in a standard tight-binding
description of the electronic structure of graphene~\cite{wallace}
one can obtain. instead of Eq.(\ref{zero}) an effective Dirac-like
Hamiltonian describing electron states near the $K$-point:
\begin{equation}
H=v_{F}\mathbf{\sigma }\left( -i \hbar \nabla
-\frac{e}{c}\mathcal{A}\right)
\end{equation}
where $v_{F}=\sqrt{3}\gamma _{0}a/2\hbar$ and $\mathcal{A}$ is the
``vector potential'' connected with the deviations of the hopping
parameters $\gamma _{i}$ from their unperturbed value $\gamma
_{0}$:
\begin{eqnarray}
\mathcal{A}_{x} &=&\frac{c}{2ev_{F}}\left( \gamma _{2}+\gamma
_{3}-2\gamma
_{1}\right),  \nonumber \\
\mathcal{A}_{y} &=&\frac{\sqrt{3}c}{2ev_{F}}\left( \gamma
_{3}-\gamma _{2}\right) ,
\end{eqnarray}
where the nearest neighbors with vectors $\left(
-a/\sqrt{3},0\right) ;\left( a/2\sqrt{3},-a/2\right) ;\left(
a/2\sqrt{3},a/2\right) $ are labelled 1,2, and 3, correspondingly.
This means that the roughness fluctuations acts on the electronic
structure near the $K$-point as an Abelian gauge
field~\cite{iordan} which is equivalent to the action of random
magnetic field. This means that the bending of graphene violates
the time-reversal symmetry for a given valley; of course, the
Umklapp processes between $K$ and $K'$ points will restore this
symmetry. As was suggested in Ref.~\cite{morozov} these effective
magnetic fields are responsible for suppression of the weak
localization effects in graphene.

Whereas smooth deformation of the graphene sheets produces gauge
field similar to electromagnetic one, different topological
defects in graphene inducing inter-valley (Umklapp) processes can
be considered as sources of a non-Abelian gauge field;
corresponding analogy with gravitation was discussed in
Refs.~\cite{crespi,vozmed}.

In relativistic quantum mechanics, chirality is a consequence of
particle-antiparticle symmetry, which also guarantees linear
energy spectrum for massless particles. The discovery of graphene
opens a unique opportunity, to investigate \textit{chiral}
particles with \textit{parabolic (non-relativistic)} energy
spectrum. Quasiparticle with such unusual properties can be found
in {\it bilayer} graphene~\cite{bilayer}. For two graphene layers,
the nearest-neighbor tight-binding approximation predicts a
gapless state with {\it parabolic} touching in $K$ and $K'$
points~\cite{bilayer,falko} (instead of conical band crossing in
graphene). The electronic spectrum in this approximation is
described by a single-particle Hamiltonian~\cite{bilayer,falko}
\begin{equation}
H=\left(
\begin{array}{cc}
0 & -\left( p_{x}-ip_{y}\right) ^{2}/2m \\
-\left( p_{x}+ip_{y}\right) ^{2}/2m & 0
\end{array}
\right)   \label{bilayer}
\end{equation}
where $p_{i}=-i\hbar \partial /\partial x_{i}$ are electron
momenta operators and $m$ is the effective mass.  Here we
neglected higher-order hopping processes which are important only
for very low Fermi energies. More accurate
consideration~\cite{peeters} gives a very small band overlap
(about 1.6 meV) but at larger energies bilayer graphene can be
treated as a gapless semiconductor. At the same time, electronic
states are still characterized by chirality and by non-trivial
Berry phase $2\pi$~\cite{bilayer,falko} (in contrast to the case
of graphene, where the Berry phase was found to be
$\pi$~\cite{kostya2,kim}).

\section{Chiral tunneling and the Klein paradox}

The term Klein paradox~\cite{klein,dombey,ktsn} usually refers to a
counterintuitive relativistic process in which an incoming electron
starts penetrating through a potential barrier if its height $V_0$
exceeds twice the electron's rest energy $mc^2$ (where $m$ is the
electron mass and $c$ the speed of light). In this case, the
transmission probability $T$ depends only weakly on the barrier
height, approaching the perfect transparency for very high barriers,
in stark contrast to the conventional, nonrelativistic tunneling
where $T$ exponentially decays with increasing $V_0$. This
relativistic effect can be attributed to the fact that a
sufficiently strong potential, being repulsive for electrons, is
attractive for positrons and results in positron states inside the
barrier, which align in energy with the electron continuum outside.
Matching between electron and positron wavefunctions across the
barrier leads to the high-probability tunneling described by the
Klein paradox.

One can think about non-relativistic quantum mechanical tunneling
though a potential barrier in terms of indeterminacy principle.
Since momentum and coordinates of a particle can not be measured
simultaneously a particle can propagate through a classically
forbidden region where the momentum is formally imaginary, and
only coordinates are well defined. In \textit{relativistic}
quantum mechanics even coordinate itself cannot be measured with
arbitrary accuracy, due to pair creation at this measurement. In
another words, the Klein paradox illustrates that the relativistic
quantum mechanics can be consistently formulated only in terms of
fields rather than individual particles~\cite{berest}.

Although Klein's gedanken experiment is now well understood, the
notion of paradox is still used widely, perhaps because the effect
has never been observed experimentally. Indeed, its observation
requires a potential drop $\approx mc^2$ over the Compton length
$\hbar/mc$, which yields enormous electric
fields~\cite{greiner,grib}(${\cal E} > 10^{16} V/cm$) and makes
the effect relevant only for such exotic situations as, for
example, positron production around super-heavy
nuclei~\cite{greiner,grib} with charge $Z \geq 170$  or
evaporation of black holes through generation of
particle-antiparticle pairs near the event horizon~\cite{page}. At
the same time, electronic structure of graphene provides us an
unique opportunity for easy experimental realization of the Klein
ultrarelativistic tunneling in $p-n$
junctions~\cite{ktsn,CF,pereira}.

Let us consider for simplicity the potential barrier that has a
rectangular shape and is infinite along the y-axis:
\begin{equation}
V\left( x\right) =\left\{
\begin{array}{cc}
V_{0}, & 0<x<D, \\
0 & \text{otherwise.}
\end{array}
\right.   \label{bar}
\end{equation}
This local potential barrier inverts charge carriers underneath
it, creating holes playing the role of positrons. For simplicity,
we assume in Eq.(\ref{bar}) infinitely sharp edges, which allows a
direct link to the case usually considered in
QED~\cite{klein,dombey}. The sharp-edge assumption is justified if
the Fermi wavelength $\lambda$ of quasiparticles is much larger
than the characteristic width of the edge smearing, which in turn
should be larger than the lattice constant (to disallow Umklapp
scattering between different valleys in graphene). Such a barrier
can be created by the electric field effect using a thin insulator
or by local chemical doping~\cite{kostya1,kostya2,kim}.
Importantly, Dirac fermions in graphene are massless and,
therefore, there is no formal theoretical requirement for the
minimal electric field ${\cal E}$ to form positron-like states
under the barrier. To create a well-defined barrier in realistic
graphene samples with a disorder, fields ${\cal E} \approx 10^{5}
V/cm$ routinely used in experiments~\cite{kostya1,kim} should be
sufficient, which is eleven orders of magnitude lower than the
fields necessary for the observation of the Klein paradox for
elementary particles.

It is straightforward to solve the tunneling problem for Dirac
electrons~\cite{ktsn}. We assume that the incident electron wave
propagates at an angle $\phi$ with respect to the $x$ axis and
then try the components of the Dirac spinor $\psi_1$ and $\psi_2$
for the Hamiltonian $H=H_{0}+ V\left(x\right)$ in the following
form:
\begin{eqnarray}
\psi _{1}\left( x,y\right)  &=&\left\{
\begin{array}{cc}
\left( e^{ik_{x}x}+re^{-ik_{x}x}\right) e^{ik_{y}y}, & x<0, \\
\left( ae^{iq_{x}x}+be^{-iq_{x}x}\right) e^{ik_{y}y}, & 0<x<D, \\
te^{ik_{x}x+ik_{y}y}, & x>D,
\end{array}
\right.   \nonumber \\
\psi _{2}\left( x,y\right)  &=& \left\{
\begin{array}{cc}
s\left( e^{ik_{x}x+i\phi }-re^{-ik_{x}x-i\phi }\right)
e^{ik_{y}y}, & x<0,
\\
s^{\prime }\left( ae^{iq_{x}x+i\theta }-be^{-iq_{x}x-i\theta
}\right)
e^{ik_{y}y}, & 0<x<D, \\
ste^{ik_{x}x+ik_{y}y+i\phi }, & x>D,
\end{array}
\right.
\end{eqnarray}
where $k_{F} = 2\pi/\lambda$ is the Fermi wavevector,
$k_{x}=k_{F}\cos \phi$ and $k_{y}=k_{F}\sin \phi$ are the
wavevector components outside the barrier, $q_{x}=\sqrt{\left(
E-V_{0}\right) ^{2}/\hbar^{2}v_{F}^{2}-k_{y}^{2}},$ $\theta =\tan
^{-1}\left(k_{y}/q_{x}\right) $ is the refraction angle,
$s=signE$, $s^{\prime}=sign\left( E-V_{0}\right)$. Requiring the
continuity of the wavefunction by matching up coefficients $a, b,
t, r$, we can find the the reflection coefficient $r$.
\begin{figure}[t]
\begin{center}\leavevmode
\includegraphics[width=0.5\linewidth]{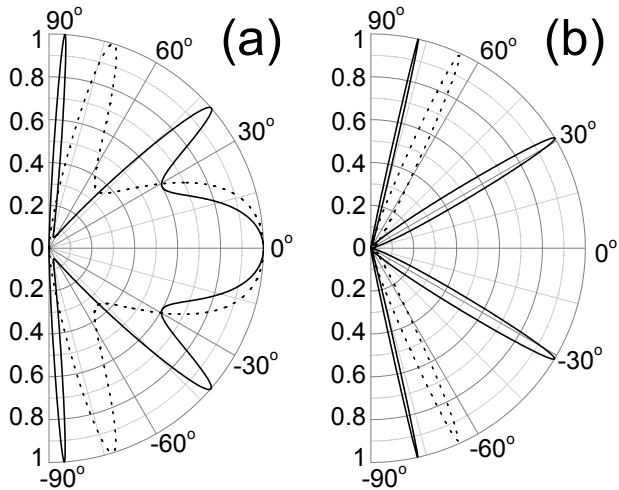}
\caption{Transmission probability $T$ through a $100$-nm-wide
barrier as a function of the incident angle for single- (a) and
bi-layer (b) graphene. The electron concentration $n$ outside the
barrier is chosen as $0.5\times 10^{12}\;$cm$^{-2}$ for all cases.
Inside the barrier, hole concentrations $p$ are $1\times 10^{12}$
and $3\times 10^{12}\;$cm$^{-2}$ for solid and dashed curves,
respectively (such concentrations are most typical in experiments
with graphene). This corresponds to the Fermi energy $E$ of incident
electrons $\approx 80$ and $17\;$meV for single- and bi-layer
graphene, respectively, and $\lambda\approx50\;$nm. The barrier
heights are (a) $200$ and (b) $50\;$meV (solid curves) and (a) $285$
and (b) $100\;$meV (dashed curves).} \label{KleinAngular}
\end{center}
\end{figure}

Fig.~\ref{KleinAngular}a shows examples of the angular dependence
of transmission probability $T = \left|
t\right|^{2}=1-\left|r\right|^{2}$ calculated using the above
expression. The barrier remains \textit{always} perfectly
transparent for angles close to the normal incidence $\phi =0$.
The latter is the feature unique to massless Dirac fermions and
directly related to the Klein paradox in QED. One can understand
this perfect tunneling in terms of the conservation of pseudospin.
Indeed, in the absence of pseudospin-flip processes (such
processes are rare as they require a short-range potential, which
would act differently on A and B sites of the graphene lattice),
an electron moving to the right can be scattered only to a
right-moving electron state or left-moving hole state. The
matching between directions of pseudospin $\sigma$ for
quasiparticles inside and outside the barrier results in perfect
tunneling. In the strictly one-dimensional case, such perfect
transmission of Dirac fermions has been discussed in the context
of electron transport in carbon nanotubes~\cite{ando,mceuen}. This
is also the case for the two-dimensional (2D) problem of graphene
which can be directly proven by consideration of the
Lippmann-Schwinger equation for scattering of arbitrary scalar
potential (see Ref.~\cite{ktsn}, supplementary information).

To elucidate which features of the anomalous tunneling in graphene
are related to the linear dispersion and which to the pseudospin
and chirality of the Dirac spectrum, it is instructive to consider
the same problem for bilayer graphene~\cite{ktsn}. Although
``massive chiral fermions'' with parabolic spectrum near the
touching point and Berry phase 2$\pi$ do not exist in the field
theory their existence in the condensed matter physics (confirmed
experimentally~\cite{bilayer}) offers a unique opportunity to
clarify the importance of chirality in the relativistic tunneling
problem described by the Klein paradox.

Charge carriers in bilayer graphene are described by an
off-diagonal Hamiltonian (\ref{bilayer}) which yields a gapless
semiconductor with chiral electrons and holes having a finite mass
$m$. An important formal difference between the tunneling problems
for single- and bi- layer graphene is that in the latter case
there are \textit{four} possible solutions for a given energy
$E=\pm \hbar^{2}k_{F}^{2}/2m$. Two of them correspond to
propagating waves and the other two to evanescent ones.
Accordingly, for constant potential $V_{i}$, eigenstates of
Hamiltonian~(\ref{bilayer}) should be written as
\begin{eqnarray}
\psi _{1}\left( x,y\right)  &=&\left(
a_{i}e^{ik_{ix}x}+b_{i}e^{-ik_{ix}x}+c_{i}e^{\kappa
_{ix}x}+d_{i}e^{-\kappa
_{ix}x}\right) e^{ik_{y}x}  \nonumber \\
\psi _{2}\left( x,y\right)  &=&s_{i}\left( a_{i}e^{ik_{ix}x+2i\phi
_{i}}+b_{i}e^{-ik_{ix}x-2i\phi _{i}}-c_{i}h_{i}e^{\kappa _{ix}x}-\frac{d_{i}%
}{h_{i}}e^{-\kappa _{ix}x}\right) e^{ik_{y}y} \label{wavebilayer}
\end{eqnarray}
where
\begin{displaymath}
s_{i}=sign\left(V_{i}-E\right); \hspace{0.5cm} \hbar
k_{ix}=\sqrt{2m\left| E-V_{i}\right| }\cos \phi _{i};
\hspace{0.5cm} \hbar k_{iy}=\sqrt{2m\left| E-V_{i}\right|}\sin
\phi _{i} $$$$ \kappa _{ix}=\sqrt{k_{ix}^{2}+2k_{iy}^{2}};
\hspace{1cm} h_{i}=\left( \sqrt{1+\sin ^{2}\phi _{i}}-\sin \phi
_{i}\right) ^{2}.
\end{displaymath}

To find the transmission coefficient through barrier (\ref{bar}),
one should set $d_{1}=0$ for $x<0,$ $b_{3}=c_{3}=0$ for $x>D$ and
satisfy the continuity conditions for both components of the
wavefunction and their derivatives. For the case of an electron
beam that is incident normally ($\phi =0$) and low barriers $V_{0}
< E$ (over-barrier transmission), we obtain $\psi_1 = - \psi_2$
both outside and inside the barrier, and the chirality of fermions
in bilayer graphene does not manifest itself. In this case,
scattering at the barrier (\ref{bar}) is the same as for electrons
described by the Schr\"{o}dinger equation. However, for any finite
$\phi$ (even in the case $V_{0} < E$), waves localized at the
barrier interfaces are essential to satisfy the boundary
conditions.

The most intriguing behavior is found for $V_{0}>E$, where
electrons outside the barrier transform into holes inside it, or
vice versa. Examples of the angular dependence of $T$ in bilayer
graphene are plotted in Fig.~\ref{KleinAngular}b. They show a
dramatic difference as compared with the case of massless Dirac
fermions. There are again pronounced transmission resonances at
some incident angles, where $T$ approaches unity. However, instead
of the perfect transmission found for normally-incident Dirac
fermions (see Fig.~\ref{KleinAngular}a), our numerical analysis
has yielded the opposite effect: Massive chiral fermions are
always perfectly reflected for angles close to $\phi =0$. At the
same time, there is always a ``magic angle'' when the transmission
probability is equal to one.

The fact that a barrier (or even a single $p-n$ junction)
incorporated in a bilayer graphene device should lead to
exponentially small tunneling current can be exploited in
developing graphene-based field effect transistors
(FET)~\cite{ktsn}. Such transistors are particularly tempting
because of their high mobility and ballistic transport at
submicron distances~\cite{kostya1,kostya2,kim}. However, the
fundamental problem along this route is that the conducting
channel in single-layer graphene cannot be pinched off due to the
Klein paradox (alternative view on this fact based on the concept
of the minimal conductivity will be discussed in the next
section). This severely limits achievable on-off ratios for such
FETs~\cite{kostya1} and, therefore, the scope for their
applications. A bilayer FET with a local gate inverting the sign
of charge carriers should yield much higher on-off ratios.

\section{Problem of minimum conductivity}

One of amazing properties of graphene is its finite minimal
conductivity which is of the order of the conductance quantum
$e^{2}/h$ per valley per spin; it is important to stress that this
is the ``quantization'' of conductivity rather than of
conductance~\cite{kostya2}. This is not only very interesting
conceptually but also important in the view of potential
applications of graphene for ballistic field-effect
transistors~\cite{kostya1,ktsn}. At the same time, this phenomenon
is intimately related with specifically quantum-relativistic
phenomenon known as Zitterbewegung~\cite{berest,zitter}.

Numerous considerations of the conductivity of a two-dimensional
massless Dirac fermion gas do give this value of the minimal
conductivity with accuracy of some factor of the order of
unity~\cite{zitter,D1,D2,ludwig,D3,shon,gorbar,D5,been,zieg}. It
is really surprising that in this case there is a final
conductivity for an \textit{ideal} crystal, that is, without any
scattering processes~\cite{ludwig,zitter,been,zieg}.

The Dirac Hamiltonian (\ref{zero}) in the secondary-quantization
representation takes the form
\begin{equation}
H=v\sum\limits_{\mathbf{p}}\Psi _{\mathbf{p}}^{\dagger }\mathbf{\sigma p}%
\Psi _{\mathbf{p}}
\end{equation}
and the corresponding expression for the current operator
\begin{equation}
\mathbf{j}=ev\sum\limits_{\mathbf{p}}\Psi_{\mathbf{p}}^{\dagger
}\mathbf{\sigma }\Psi_{\mathbf{p}}
\end{equation}
where $\mathbf{p}$ is the momentum and $\Psi
_{\mathbf{p}}^{\dagger }=\left( \psi _{\mathbf{p}1}^{\dagger
},\psi _{\mathbf{p}2}^{\dagger }\right) $  are pseudospinor
electron operators. Here we omit spin and valley indices (so,
keeping in mind applications to graphene, the results for the
conductivity should be multiplied by 4 due to two spin projections
and two valleys). Straightforward calculations result in the
following time evolution of the electron annihilation operator
\begin{equation}
\Psi _{\mathbf{p}}\left( t\right) =\frac{1}{2}\left[ e^{-i\epsilon _{\mathbf{%
p}}t}\left( 1+\frac{\mathbf{p\sigma }}{p}\right) +e^{i\epsilon _{\mathbf{p}%
}t}\left( 1-\frac{\mathbf{p\sigma }}{p}\right) \right] \Psi
_{\mathbf{p}}  \label{second}
\end{equation}
and for the current operator
\begin{eqnarray}
\mathbf{j}\left( t\right)  &=&\mathbf{j}_{0}\left( t\right) +\mathbf{j}%
_{1}\left( t\right) +\mathbf{j}_{1}^{\dagger }\left( t\right)   \nonumber \\
\mathbf{j}_{0}\left( t\right)  &=&ev\sum\limits_{\mathbf{p}}\Psi _{\mathbf{p}%
}^{\dagger }\frac{\mathbf{p}\left( \mathbf{p\sigma }\right) }{p^{2}}\Psi _{%
\mathbf{p}}  \nonumber \\
\mathbf{j}_{1}\left( t\right)  &=&\frac{ev}{2}\sum\limits_{\mathbf{p}}\Psi _{%
\mathbf{p}}^{\dagger }\left[ \mathbf{\sigma }-\frac{\mathbf{p}\left( \mathbf{%
p\sigma }\right) }{p^{2}}+\frac{i}{p}\mathbf{\sigma \times p}\right] \Psi _{%
\mathbf{p}}e^{2i\epsilon _{\mathbf{p}}t}  \label{current}
\end{eqnarray}
where $\epsilon _{\mathbf{p}}=vp/\hbar$ is the particle frequency.
The last term in Eq.(\ref{current}) corresponds to the
``Zitterbewegung'', a phenomenon connected with the uncertainty of
the position of relativistic quantum particles due to the
inevitable creation of particle-antiparticle pairs at the position
measurement~\cite{berest}.

To calculate the conductivity $\sigma \left( \omega \right)$
following Ref.~\cite{zitter} we will try first to use the Kubo
formula~\cite{zubarev} which reads for two-dimensional isotropic
case:
\begin{equation}
\sigma \left( \omega \right) =\frac{1}{2A}\int\limits_{0}^{\infty
}dte^{i\omega t}\int\limits_{0}^{\beta }d\lambda \left\langle \mathbf{j}%
\left( t-i\lambda \right) \mathbf{j}\right\rangle  \label{kubo11}
\end{equation}
where $\beta =T^{-1}$ is the inverse temperature, $A$ is the
sample area. In the static limit $\omega =0$ taking into account
the Onsager relations and analyticity of the correlators
$\left\langle \mathbf{j}\left( z\right) \mathbf{j}\right\rangle$
for $- \beta < \mathrm{Im}z \leq 0 $ one has~\cite{zubarev}
\begin{equation}
\sigma =\frac{\beta }{4A}\int\limits_{-\infty }^{\infty
}dt\left\langle \mathbf{j}\left( t\right) \mathbf{j}\right\rangle
.  \label{kubo}
\end{equation}
Usually, for ideal crystals, the current operator commutes with
the Hamiltonian and thus $\mathbf{j}\left( t\right)$ does not
depend on time. In that case, due to Eq.(\ref{kubo11}) the
frequency-dependent conductivity contains only the Drude peak
\begin{equation}
\sigma _{D}\left( \omega \right) =\frac{\pi
}{2A}\lim_{T\rightarrow 0}\frac{\left\langle
\mathbf{j}^{2}\right\rangle }{T}\delta \left( \omega \right)
\label{Drude}
\end{equation}
Either the spectral weight of the Drude peak is finite and, thus,
the static conductivity is infinite, or it is equal to zero. It is
easy to check that for the system under consideration the spectral
weight of the Drude peak is proportional to the modulus of the
chemical potential $\left| \mu \right| $ (cf. Eq.(44) of
Ref.~\cite{D5}) and thus vanishes at zero doping ($\mu =0$). It is
the Zitterbewegung, i.e. the oscillating term
$\mathbf{j}_{1}\left( t\right) $ which is responsible for
nontrivial behavior of the conductivity for zero temperature and
zero chemical potential (that is, the case of no charge carriers).
A straightforward calculation gives a formal result
\begin{equation}
\sigma =\frac{\pi e^{2}}{2h}\int\limits_{0}^{\infty }d\epsilon
\epsilon \delta ^{2}\left( \epsilon \right)  \label{sigma}
\end{equation}
where one delta-function originates from the integration over $t$
in Eq.(\ref {kubo}) and the second one - from the derivative of
the Fermi distribution function appearing at the calculation of
the average over product of Fermi-operators. Of course, the square
of the delta function is not a well-defined object and thus
Eq.(\ref{sigma}) is meaningless before specification of the way
how one should regularize the delta-functions. After
regularization the integral in Eq.(\ref{sigma}) is finite, but its
value depends on the regularization procedure (see
Refs.~\cite{ludwig,zieg}).

\begin{figure}[tbp]
\includegraphics[width=9cm]{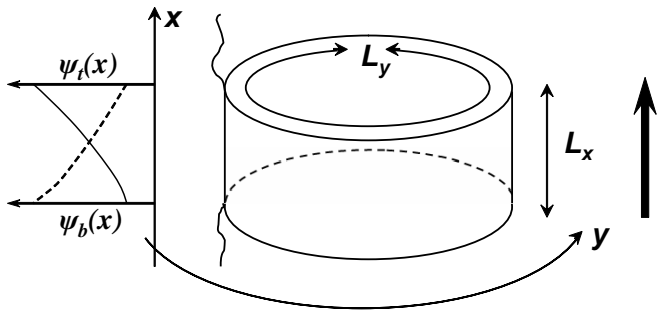}
\caption{Geometry of the sample. The thick arrow shows the
direction of current. $\protect\psi_t$ (solid line) and
$\protect\psi_b$ (dashed line) are wave functions of the edge
states localized near the top and the bottom of the sample,
correspondingly. } \label{cylindr}
\end{figure}

To circumvent the problem of ambiguity in the expression for
$\sigma$ in Eq.(\ref{sigma}) it is instructive to follow the
alternative Landauer approach to calculate the
conductivity~\cite{beenakker,buttiker}. Let us assume that our
sample is the ring of length $L_{y}$ in $y$ direction; we will use
the Landauer formula to calculate the conductance in the $x$
direction (see Fig. \ref{cylindr}). There is still an uncertainty
in the sense that the conductivity turns out to be dependent on
the shape of the sample. To have a final transparency we should
keep $L_{x}$ finite. On the other hand, periodic boundary
conditions in $y$ direction are nonphysical and we have to choose
$L_{y}$ as large as possible to weaken their effects. Thus, for
two-dimensional situation one should choose $L_{x}\ll L_{y}.$

In the coordinate representation the Dirac equation at zero energy
takes the form
\begin{eqnarray}
(K_{x}+iK_{y})\psi _{1} &=&0 \nonumber \\
(K_{x}-iK_{y})\psi _{2} &=&0
\end{eqnarray}
where $K_{i}=-i\frac{\partial }{\partial x_{i}}.$ General
solutions of these equations are just arbitrary analytical (or
complex conjugated analytical) functions:
\begin{eqnarray}
\psi _{1} &=&\psi _{1}\left( x+iy\right) ,  \nonumber \\
\psi _{2} &=&\psi _{2}\left( x-iy\right) .
\end{eqnarray}
Due to periodicity in the $y$ direction both wave functions should
be proportional to $\exp \left( ik_{y}y\right) $ where $k_{y}=2\pi
n/L_{y},n=0,\pm 1,\pm 2,....$ This means that the dependence on
$x$ is also fixed: the wave functions are proportional to $\exp
\left( \pm 2\pi nx/L_{y}\right) .$ They correspond to the states
localized near the bottom and the top of the sample (see Fig.
\ref{cylindr}).

To use the Landauer formula, we should introduce boundary
conditions at the sample edges ($x=0$ and $x=L_{x}$). To be
specific, let us assume that the leads are made of doped graphene
with the potential $V_{0}<0$ and the Fermi energy
$E_{F}=vk_{F}=-V_{0}.$ The wave functions in the leads are
supposed to have the same $y$-dependence, that is, $\psi
_{1,2}\left( x,y\right) =\psi _{1,2}\left( x\right) \exp \left(
ik_{y}y\right)$. Thus, one can try the solution of the Dirac
equation in the following form:
\begin{eqnarray}
\psi _{1}\left( x\right)  &=&\left\{
\begin{array}{cc}
e^{ik_{x}x}+re^{-ik_{x}x}, & x<0 \\
ae^{k_{y}x}, & 0<x<L_{x} \\
te^{ik_{x}x}, & x>L_{x}
\end{array}
\right.   \nonumber \\
\psi _{2}\left( x\right)  &=&\left\{
\begin{array}{cc}
e^{ik_{x}x+i\phi }-re^{-ik_{x}x-i\phi }, & x<0 \\
be^{-k_{y}x}, & 0<x<L_{x} \\
te^{ik_{x}x+i\phi }, & x>L_{x}
\end{array}
\right.   \label{solution}
\end{eqnarray}
where $\sin \phi =k_{y}/k_{F},k_{x}=\sqrt{k_{F}^{2}-k_{y}^{2}}.$
>From the conditions of the continuity of the wave functions, one
can find the transmission coefficient
\begin{equation}
T_{n}=\left| t\left( k_{y}\right) \right| ^{2}=\frac{\cos ^{2}\phi
}{\cosh ^{2}(k_{y}L_{x})-\sin ^{2}\phi }.  \label{T1}
\end{equation}
Furthermore, one should assume that $k_{F}L_{x}\gg 1$ and put
$\phi \simeq 0$ in Eq.(\ref{T1}). Thus, the trace of the
transparency which is just the conductance (in units of $e^{2}/h$)
is
\begin{equation}
TrT=\sum\limits_{n=-\infty }^{\infty }\frac{1}{\cosh
^{2}(k_{y}L_{x})}\simeq \frac{L_{y}}{\pi L_{x}}.  \label{T2}
\end{equation}
Assuming that the conductance is equal to $\sigma
\frac{L_{y}}{L_{x}}$ one finds the contribution to the
conductivity equal to $e^{2}/(\pi h)$.
Experimentally~\cite{kostya2}, it is close to $e^{2}/h$, that is,
roughly, three times larger than this estimation. This discrepancy
known as a ``missed pi(e) problem'' is still not solved.

One should stress that this is the gapless character of the
spectrum and the chirality rather than the linear dispersion
relation which lead to the minimal conductivity. Different
estimations for the bilayer graphene give the same order of
magnitude for the
conductivity~\cite{zitter,zitter2,ando2,nils,cserti,been2}, in
agreement with experimenatl observations~\cite{bilayer}.

Another approach to qualitative understanding of the minimal
conductivity is based on the Klein paradox~\cite{ktsn}. In
conventional 2D systems, strong enough disorder results in
electronic states that are separated by barriers with
exponentially small transparency~\cite{ziman,lifshitz}. This is
known to lead to the Anderson localization. In contrast, in both
graphene materials all potential barriers are relatively
transparent ($T\approx1$ at least for some angles) which does not
allow charge carriers to be confined by potential barriers that
are smooth on an atomic scale. Therefore, different electron and
hole ``puddles'' induced by disorder are not isolated but
effectively percolate, thereby suppressing localization. In the
absence of the localization, a standard Mott estimation of the
minimal conductivity assuming that the mean-free path cannot be
smaller that  electron wave-length~\cite{mott} immediately gives
the minimal conductivity $\approx e^2/h$~\cite{private}.

\section{Index theorem and anomalous quantum Hall effect}

The anomalous QHE in graphene~\cite{kostya2,kim} is probably the
most striking demonstration of the massless character of the
charge carrier spectrum in graphene. In two-dimensional systems
with a constant magnetic field ${\bf B}$ perpendicular to the
system plane the energy spectrum is discrete (Landau
quantization). For the case of massless Dirac fermions the energy
spectrum takes the form~\cite{mcclure,GS}
\begin{equation}
E_{\nu \sigma }=\sqrt{2\left| e\right| B\hbar v_{F}^{2}\left( \nu
+1/2\pm 1/2\right) } \label{landaulevel}
\end{equation}
where $v_{F}$ is the electron velocity, $\nu = 0,1,2...$ is the
quantum number and the term with $\pm 1/2$ is connected to the
chirality. Just to remind that in the usual case of the parabolic
dispersion relation the Landau level sequence is
$E=\hbar\omega_c(\nu +1/2)$ where $\omega_c$ is the frequency of
electron rotation in magnetic field (cyclotron
frequency)~\cite{Ashcroft,VK}.

An important peculiarity of Landau levels for massless Dirac
fermions is the existence of zero-energy states (with $\nu =0$ and
the minus sign in equation (\ref{landaulevel})). This situation
differs fundamentally from usual semiconductors with parabolic
bands where the first Landau level is shifted by
$\hbar\omega_c/2$. The existence of the zero-energy Landau level
leads to an anomalous QHE with \textit{half-integer} quantization
of the Hall conductivity, instead of the \textit{integer} one (for
a review of the QHE, see, e.g., Ref.~\cite{QHE}). Usually, all
Landau levels have the same degeneracy (a number of electron
states with a given energy) which is just proportional to a
magnetic flux through the system. As a result, plateaus in the
Hall conductivity corresponding to the filling of the first $\nu$
levels are just integer (in units of the conductance quant
$e^{2}/h$). For the case of massless Dirac electrons, the
zero-energy Landau level has twice smaller degeneracy than any
other level (it corresponds to the minus sign in the equation (1)
whereas the energy level proportional to $\sqrt{p}$ with integer
$p\geq 1$ are obtained twice, for $\nu =p$ and minus sign, and for
$\nu =p-1$ and plus sign). A discovery of this ``half-integer
QHE''~\cite{kostya2,kim} was the most direct evidence of the Dirac
fermions in graphene.

A deeper view on the origin of the zero-energy Landau level and
thus the anomalous QHE is provided by the famous Atiyah-Singer
index theorem which plays an important role in the modern quantum
field theory and theory of superstrings~\cite{index}. The Dirac
equation has a charge-conjugation symmetry between electrons and
holes. This means that for any electron state with a positive
energy $E$ a corresponding conjugated hole state with the energy
$-E$ should exist. However, the states with zero energy can be, in
general, anomalous. For curved space (e.g., for a deformed
graphene sheet with some defects of crystal structure) and/or in
the presence of so called ``gauge fields'' (the usual
electromagnetic field provides just the simplest example of these
fields) sometimes an existence of states with zero energy is
guaranteed by topological reasons, this states being chiral (in
our case this means that depending on the sign of the magnetic
field there is only sublattice A or sublattice B states which
contribute to the zero-energy Landau level). This means, in
particular, that the number of these states expressed in terms of
total magnetic flux is a topological invariant and remains the
same even if the magnetic field is inhomogeneous~\cite{kostya2}.
This is an important conclusion since, as discussed in section 1,
the ripples on graphene create strong effective inhomogeneous
magnetic fields leading, in particular, to suppression of weak
localization~\cite{morozov}. However, due to these topological
arguments they cannot destroy the anomalous QHE in graphene. About
applications of the index theorem to two-dimensional systems, and,
in particular, to graphene see also Refs.~\cite{annphys,stone}.

An alternative view on the origin of the anomalous QHE in graphene
is based on the concept of the ``Berry phase'' which was discussed
above in connection with the Klein paradox. When chiral electron
moves along the close contour its wave function gains an
additional phase $\phi = \pi$. In quasiclassical terms (see, e.g.,
Refs.~\cite{VK,mikitik}), stationary states are nothing but
electron standing waves and they can exist if the electron orbit
contains, at least, half of the wavelength. Due to the additional
phase shift by the Berry phase, this condition is satisfied
already for the zeroth length of the orbit, that is, for zero
energy. Other aspects of the QHE in graphene are considered in
papers~\cite{levitov,gus,per,cas}.

In the previous section we have discussed an absence of
localization in graphene as one of the main consequences of
relativistic quantum effects such as the Klein paradox. At the
same time, the localization is of crucial importance for the QHE
providing a well-defined Hall plateau~\cite{QHE}. Here we
demonstrate that the localization by a scalar potential $V$ in the
magnetic field is possible. For simplicity, let us assume the case
of a weak one-dimensional inhomogeneity with the potential
$V=V(y)$ much smaller than the cyclotron quantum of the order of
$\hbar v_F /l$ where $l = \left( \hbar c/ |e| B \right)^{1/2} $ is
the magnetic length.

Using a standard gauge for the vector potential $A_x
=A_z=0,A_y=Bx$ one can write the Dirac Hamiltonian for the problem
as
\begin{equation}
\widehat{H}=\widehat{H}_{1}+\widehat{H}_{2}  \label{hall1}
\end{equation}
where
\begin{equation}
\widehat{H}_{1}=\left(
\begin{array}{cc}
0 & -i\hbar v_{F}\left( \frac{\partial }{\partial
x}-\frac{x}{l^{2}}\right)
\\
-i\hbar v_{F}\left( \frac{\partial }{\partial
x}+\frac{x}{l^{2}}\right)  & 0
\end{array}
\right) ,  \label{hall2}
\end{equation}
\begin{equation}
\widehat{H}_{2}=\left(
\begin{array}{cc}
V\left( y\right)  & \hbar v_{F}\frac{\partial }{\partial y} \\
-\hbar v_{F}\frac{\partial }{\partial y} & V\left( y\right)
\end{array}
\right) .  \label{hall3}
\end{equation}
Let us try eigenfunctions of the Hamiltonian (\ref{hall1}) with
the energy $E $ as an expansion in the basis of appropriate
harmonic oscillator problem:
\begin{eqnarray}
\psi _{i} &=&\sum\limits_{\nu =0}^{\infty }\int \frac{dk_{y}}{2\pi
}c_{\nu }^{(i)}\left( k_{y}\right) e^{ik_{y}y}\varphi _{\nu
}\left( k_{y},x\right) ,
\nonumber \\
\varphi _{\nu }\left( k_{y},x\right)  &=&D_{\nu }\left(
\frac{\sqrt{2}\left( x-l^{2}k_{y}\right) }{l}\right) ,
\label{hall4}
\end{eqnarray}
where $i=1,2$ is the pseudospinor index and $D_{\nu }\left(
z\right) \sim \exp \left( -z^{2}/4\right) H_{\nu }\left(
z/\sqrt{2}\right) $ the Weber functions~\cite{whittaker}. After
straightforward calculations one obtains a set of equations for
the expansion coefficients $c_{\nu }^{(i)}\left( k_{y}\right) $:
\begin{eqnarray}
-\frac{\sqrt{2}}{l}\left( 1-\delta _{\nu ,0}\right) c_{\nu
}^{(2)}\left(
k_{y}\right)  &=&\frac{iE}{\hbar v_{F}}c_{\nu }^{(1)}\left( k_{y}\right) -%
\frac{i}{\hbar v_{F}}\sum\limits_{\nu ^{\prime }=0}^{\infty }\int \frac{%
dq_{y}}{2\pi }v\left( k_{y}-q_{y}\right) c_{\nu ^{\prime
}}^{(1)}\left( q_{y}\right) \left\langle \nu ,k_{y}|\nu ^{\prime
},q_{y}\right\rangle ,
\nonumber \\
\frac{\sqrt{2}}{l}\left( 1+\nu \right) c_{\nu }^{(1)}\left( k_{y}\right)  &=&%
\frac{iE}{\hbar v_{F}}c_{\nu }^{(2)}\left( k_{y}\right) -\frac{i}{\hbar v_{F}%
}\sum\limits_{\nu ^{\prime }=0}^{\infty }\int \frac{dq_{y}}{2\pi
}v\left( k_{y}-q_{y}\right) c_{\nu ^{\prime }}^{(2)}\left(
q_{y}\right) \left\langle
\nu ,k_{y}|\nu ^{\prime },q_{y}\right\rangle ,  \nonumber \\
\left\langle \nu ,k_{y}|\nu ^{\prime },q_{y}\right\rangle  &=&\int
dx\varphi _{\nu }\left( k_{y},x\right) \varphi _{\nu ^{\prime
}}\left( q_{y},x\right) , \label{hall5}
\end{eqnarray}
$v(q)$ is a Fourier component of $V(y).$ For the case of a weak
($\left| V\left( y\right) \right| \ll \hbar v_{F}/l$) and smooth
potential one can neglect a mixing of different Landau bands,
similar to the case of QHE for conventional electron gas
\cite{QHE}. As a result of direct calculations one can demonstrate
that in this case the solution of Eq.(\ref{hall5}) describes the
Landau wave functions of nonperturbed problem with the orbit
center $y_{0}$ satisfying the equation
\begin{equation}
E\pm \frac{\hbar v_{F}}{l}\sqrt{2\nu }=V\left( y_{0}\right).
\label{hall6}
\end{equation}
Thus, the Landau levels in the case under consideration are just
smeared into the bands of localized states with the localization
radius of order of the magnetic length, similar to the case of the
usual QHE~\cite{QHE}.

\section{Nonlinear screening of charge impurities}

In QED, vacuum polarization effects due to creation of virtual
electron-positron pairs by an external potential are of a crucial
importance~\cite{blp,migdal}. Due to the smallness of the fine
structure constant $\alpha_{QED} = e^2 /\hbar c \simeq 1/137$
these effects leading to logarithmic corrections to the Coulomb
potential are rather small, except the case of very small
distances. In graphene, the effective fine structure constant
$\alpha = e^2 / \hbar v_F \epsilon$ is of the order of unity
($\epsilon$ is the dielectric constant due to substrate; for the
case of graphene on quartz one should choose~\cite{ando11}
$\epsilon \simeq 2.4-2.5$) and the corresponding effects can be of
crucial importance for transport properties~\cite{nonlinear}. Due
to the absence of explicitely small parameters describing
correlation effects in graphene it is very difficult to consider
this problem rigorously and the situation still is controversial.
Nevertheless, the vacuum polarization effects are in our opinion
important and should be taken into account in any future theory.

A general nonlinear theory of screening in the system of
interacting particles can be formulated in a framework of the
density functional approach~\cite{kohn}. In this theory a total
potential $V\left( \mathbf{r}\right)$ acting on electrons is equal
\begin{equation}
V\left( \mathbf{r}\right) =V_{0}\left( \mathbf{r}\right)
+V_{ind}\left( \mathbf{r}\right)   \label{total}
\end{equation}
where $V_{0}\left( \mathbf{r}\right)$ is an external potential and
$V_{ind}\left( \mathbf{r}\right)$ is a potential induced by a
redistribution of electron density:
\begin{equation}
V_{ind}\left( \mathbf{r}\right) =\frac{e^{2}}{\epsilon }\int d\mathbf{r}%
^{\prime }\frac{n\left( \mathbf{r}^{\prime }\right)
-\overline{n}}{\left| \mathbf{r-r}^{\prime }\right| }+V_{xc}\left(
\mathbf{r}\right),  \label{DF}
\end{equation}
where the first term is the Hartree potential and the second one
is the exchange-correlation potential. We consider here only a
redistribution of charge carriers in the external impurity
potential
\begin{equation}
V_{0}\left( \mathbf{r}\right) =\frac{Ze^{2}}{\epsilon r}
\label{impur}.
\end{equation}
taking into account contributions of the crystal lattice potential
and of electrons in completely filled bands via a dielectric
constant $\epsilon$ and compensated homogeneous charge density
$-e\overline{n}$. Here $Z$ is the dimensionless impurity charge
(to be specific, we will assume $Z>0$; it can be easily
demonstrated that, actually, in our final expressions $Z$ should
be just replaced by $\left| Z\right| $). This kind of approach is
valid at a spacial scale much larger than a lattice constant; in
all other aspects, it is formally exact until we specify the
expressions for $V_{xc}$ and $n\left[V \left(\mathbf{r} \right)
\right]$.

The Thomas-Fermi theory~\cite{lieb} which is, actually, the
simplest approximation in the density functional approach, was
used in Ref.~\cite{nonlinear}. It is based on the two assumptions:
(i) we neglect the exchange-correlation potential in comparison
with the Hartree potential in Eq.(\ref{DF}) and (ii) we put
$n\left( \mathbf{r}^{\prime }\right) = n\left[ \mu -V\left(
\mathbf{r}^{\prime }\right) \right]$, where $n\left( \mu \right) $
is a density of a homogeneous electron gas with chemical potential
$\mu$. The former assumption means that we are interested in the
long-wavelength response of the electron system and thus the
long-range Coulomb forces dominate over the local
exchange-correlation effects. The latter one holds provided that
the external potential is smooth enough. A rigorous statement is
that an addition of a \textit{constant} potential is equivalent to
the shift of the chemical potential. In particular, the
Thomas-Fermi theory gives an exact expression for a static
inhomogeneous dielectric function $\epsilon(q)$ in the limit of
small wavevectors $q \rightarrow 0$\cite{VK}. The Thomas-Fermi
theory of atoms is asymptotically exact in the limit of infinite
nuclear charge~\cite{lieb}.

In the Thomas-Fermi theory Eq.(\ref{DF}) reads
\begin{equation}
V_{ind}\left( \mathbf{r}\right) = \frac{e^{2}}{\epsilon }\int d\mathbf{r}%
^{\prime }\frac{n\left[ \mu -V\left( \mathbf{r}^{\prime }\right)
\right] -n\left( \mu \right) }{\left| \mathbf{r-r}^{\prime
}\right| }.  \label{ind}
\end{equation}

The function $n\left( \mu \right) $ is expressed via the density
of states $N\left( E\right) $:
\begin{equation}
n\left( \mu \right) =\int dEf\left( E\right) N\left( E\right)
=\int\limits^{\mu }dEN\left( E\right)   \label{nmu}
\end{equation}
where $f\left( E\right)$ is the Fermi function, and the last
equality is valid for zero temperature (we will restrict ourselves
here only to this case). For the case of graphene with the linear
energy spectrum near the crossing points $K$ and $K^{\prime }$ one
has
\begin{equation}
n\left( \mu \right) =\frac{1}{\pi }\frac{\mu \left| \mu
\right|}{\hbar ^{2}v_{F}^{2}}, \label{nmu1}
\end{equation}
where we have taken into account a factor 4 due to two valleys and
two spin projections.

Let us start first with the case of zero doping ($\mu =0$) where,
according to the linear response theory, there is no screening at
all. Substituting Eqs.(\ref{ind}), (\ref{impur}), and (\ref{nmu1})
into Eq.(\ref{total}), introducing the notation
\begin{equation}
V\left( r\right) =\frac{e^{2}}{\epsilon r}F\left( r\right)
\label{F}
\end{equation}
and integrating over the polar angle of vector $\mathbf{r}^{\prime
}$, we obtain a nonlinear integral equation for the function
$F\left( r\right):$
\begin{equation}
F\left( r\right) =Z-\frac{2Q}{\pi }\int\limits_{0}^{\infty
}\frac{dr^{\prime
}}{r^{\prime }}\frac{r}{r+r^{\prime }}K\left( \frac{2\sqrt{rr^{\prime }}}{%
r+r^{\prime }}\right) F^{2}\left( r^{\prime }\right)
\label{integral}
\end{equation}
where
\begin{equation}
K\left( k\right) =\int\limits_{0}^{\pi /2}\frac{d\varphi
}{\sqrt{1-k^{2}\sin ^{2}\varphi }}  \label{K}
\end{equation}
is the complete elliptic integral,
\begin{equation}
Q=2\left( \frac{e^{2}}{\epsilon \hbar v_{F}}\right) ^{2};
\label{Q}
\end{equation}
for the case of graphene on SiO$_{2}$ $Q\simeq 2$.

An approximate solution of the integral equation (\ref{integral})
at $r \gg a$ gives the following result~\cite{nonlinear}
\begin{equation}
F\left( r\right) \simeq \frac{Z}{1+ZQ\ln \frac{r}{a}}
\label{solution}
\end{equation}
which is formally similar to that in QED~\cite{migdal} but with a
much larger value of the interaction constant. Note that
logarithmically divergent corrections to the electron effective
mass in undoped graphene due to the vacuum polarization effects
have been considered in Ref.~\cite{gonzales}.

For the case of doped graphene the main result of
Ref.~\cite{nonlinear} is the replacement of the bare charge
impurity $Z$ by
\begin{equation}
Z^{\ast }\simeq \frac{Z}{1+ZQ\ln \frac{1}{\kappa a}}.
\label{final}
\end{equation}
where
\begin{equation} \kappa =\frac{4e^{2} \left| \mu
\right|}{\epsilon \hbar ^{2}v_{F}^{2}} \label{kappa}
\end{equation}
is the inverse screening radius. This weakens essentially this
scattering mechanism since $Q\ln \frac{1}{\kappa a}$ is of order
of ten for typical charge carrier concentrations. Perturbative
estimations of the electron mobility~\cite{nomura} should be thus
multiplied by this factor squared. As a result, the mobility for
the same parameters turns out to be two order of magnitude larger.
Instead of a concentration-independent mobility, we obtain a
mobility proportional to $\ln^{2}\left( k_{F}a\right)$. More
accurately, one should use an expression for the mobility obtained
by Ando\cite{ando11} (see Eq.(3.27) and Fig. 5 of that paper) but
with the replacement of $Z$ by $Z^{\ast }$ when calculating the
strength of the Coulomb interaction.

\section{Scattering of Dirac fermions by short-range impurity potential}

Here we will discuss quantum relativistic effects in the electrons
scattering by a short-range potential. It appears that in the case
of graphene the contribution of such defects to the resistivity is
essentially smaller than for the conventional nonrelativistic
two-dimensional electron gas. We argue that this conclusion should
help our understanding of remarkably high charge carrier
mobilities observed in graphene~\cite{kostya1,kostya2,kim}.

Let us consider the case of a small concentration of point defects
(to be specific, we will call them impurities) with the
concentration $n_{imp}$ and the angle-dependent scattering
cross-section $\sigma \left( \phi \right) .$ Then the defect
contribution to the resistivity $\rho $ reads~\cite{shon}
\begin{eqnarray}
\rho  &=&\frac{2}{e^{2}v_{F}^{2}N\left( E_{F}\right)
}\frac{1}{\tau \left(
k_{F}\right) },  \nonumber \\
\frac{1}{\tau \left( k_{F}\right) }
&=&n_{imp}v_{F}\int\limits_{0}^{2\pi }d\phi \frac{d\sigma \left(
\phi \right) }{d\phi }\left( 1-\cos \phi \right) \label{resist}
\end{eqnarray}
where $N\left( E_{F}\right) =2k_{F}/\pi \hbar v_{F}$ is the
density of states at the Fermi level (taking into account the
double spin degeneracy and two valleys), $\tau $ is the mean-free
path. Note that the product $v_{F}N\left( E_{F}\right) $ is
proportional to $k_{F}=\sqrt{\pi n}$ ($n$ is the electron
concentration) for both ultrarelativistic and nonrelativistic
two-dimensional electron gas and thus any essential difference in
their transport properties can be related only to the scattering
cross-section.

To determine the scattering cross section one has to solve the
two-dimensional Dirac equation which, for the case of massless
particles and radially symmetric scattering potential $V\left(
r\right) $ takes the from (cf. Ref.~\cite{berest} for
three-dimensional case):
\begin{eqnarray}
\frac{dg_{l}\left( r\right) }{dr}-\frac{l}{r}g_{l}\left( r\right) -\frac{i}{%
\hbar v_{F}}\left[ E-V\left( r\right) \right] f_{l}\left( r\right)
&=&0,
\nonumber \\
\frac{df_{l}\left( r\right) }{dr}+\frac{l+1}{r}f_{l}\left( r\right) -\frac{i%
}{\hbar v_{F}}\left[ E-V\left( r\right) \right] g_{l}\left(
r\right)  &=&0, \label{radial}
\end{eqnarray}
where $l=0,\pm 1,...$ is the angular-momentum quantum number,
$g_{l}\left( r\right) e^{il\phi }$ and $f_{l}\left( r\right)
e^{i(l+1)\phi }$ are components of the Dirac pseudospinor; to be
specific we will consider the case of electrons $E=\hbar
v_{F}k>0$.

Modifying a standard scattering theory~\cite{newton} for the
two-dimensional case one should try the solutions of
Eq.(\ref{radial}) \ outside the region of action of the potential
in the form
\begin{eqnarray}
g_{l}\left( r\right)  &=&A\left[ J_{l}\left( kr\right)
+t_{l}H_{l}^{(1)}\left( kr\right) \right] ,  \nonumber \\
f_{l}\left( r\right)  &=&iA\left[ J_{l+1}\left( kr\right)
+t_{l}H_{l+1}^{(1)}\left( kr\right) \right] ,  \label{bessel}
\end{eqnarray}
where the terms proportional to Bessel (Hankel) functions describe
incident (scattering) waves,
\begin{equation}
\frac{d\sigma \left( \phi \right) }{d\phi }=\frac{2}{\pi k}\left|
\sum\limits_{l=-\infty }^{\infty }t_{l}e^{il\phi }\right| ^{2}.
\label{crosssect}
\end{equation}

The Dirac equation for ultrarelativstic particles (\ref{radial})
has as important symmetry with respect to replacement
$f\longleftrightarrow g,l\longleftrightarrow -l-1$ which means
$t_{l}=t_{-l-1}.$ Thus, Eq.(\ref {crosssect}) can be rewritten in
the form
\begin{equation}
\frac{d\sigma \left( \phi \right) }{d\phi }=\frac{2}{\pi k}\left|
\sum\limits_{l=0}^{\infty }t_{l}\cos \left[ \left( l+1/2\right)
\phi \right] \right| ^{2}.  \label{crosssect1}
\end{equation}
The back scattering ($\phi =\pi $) is absent rigorously, as was
discussed  in the section 2.

For the simplest case of the potential $V\left( r\right) =V_{0}$
at $r<R$ and $V\left( r\right) =0$ at $r>R$, using boundary
conditions of continuity of the wave functions at $r=R$ one finds
\begin{equation}
t_{l}\left( k\right) =\frac{J_{l}\left( qR\right) J_{l+1}\left(
kR\right) -J_{l}\left( kR\right) J_{l+1}\left( qR\right)
}{H_{l}^{(1)}\left( kR\right) J_{l+1}\left( qR\right) -J_{l}\left(
qR\right) H_{l+1}^{(1)}\left( kR\right) }  \label{tt}
\end{equation}
where $q=\left( E-V_{0}\right) /\hbar v_{F}$. For the case of small energy, $%
E\ll V_{0},kR\ll 1$ which is typical for graphene one has
\begin{equation}
t_{l}\left( k\right) \simeq \frac{\pi i}{\left( l!\right) ^{2}}\frac{%
J_{l+1}\left( qR\right) }{J_{l}\left( qR\right) }\left(
\frac{kR}{2}\right)^{2l +1}  \label{tt1}
\end{equation}
and thus the $s$-scattering ($l=0$) dominates. Substituting
Eq.(\ref{tt1}) into Eqs.(\ref{crosssect1}) and (\ref{resist}), one
finds the estimation for the impurity contribution to the
resistivity $\rho \simeq \left( h/4e^{2}\right) n_{imp}R^{2}.$
This means that scattering centers with the radius of potential
$R$ of order of interatomic distances and small concentration are
irrelevant for the electron transport in graphene giving
negligible contribution to the resistivity.

If we would repeat the same calculations for nonrelativistic
electron gas~\cite{adhikari} we would obtain, instead of
Eq.(\ref{tt})
\begin{equation}
t_{l}\left( k\right) =\frac{\left( k/q\right) J_{l}\left(
qR\right)
J_{l+1}\left( kR\right) -J_{l}\left( kR\right) J_{l+1}\left( qR\right) }{%
H_{l}^{(1)}\left( kR\right) J_{l+1}\left( qR\right) -\left(
k/q\right) J_{l}\left( qR\right) H_{l+1}^{(1)}\left( kR\right) }
\label{ttnonrel}
\end{equation}
where $k$ and $q$ are, again, wavevectors outside and inside the
potential region. In this case the $s$-scattering phase vanishes for
small energies not linearly but only logarithmically and one obtains
much larger resistivity
\begin{equation}
\rho \simeq \frac{h}{4e^{2}}\frac{n_{imp}}{n\ln ^{2}\left(
k_{F}R\right)}  \label{resistnonrel}
\end{equation}
(note that the same estimation holds for the case of massless
Dirac electrons in a particular case of the resonant scattering
when $J_{0}\left( qR\right) =0$).

So, the basic conclusion is that whenever any potential works as a
resonant scatterer in the case of massive particles, the
scattering is non-resonant in the case of massless Dirac fermions
in graphene, except some special values of parameters.

\section*{Conclusions}

The examples considered demonstrate that graphene provides an
unexpected bridge between the condensed matter physics and quantum
field theory. The impact of the experimental discovery of this
material on different areas of physics and industry can't be
overestimated. First of all, graphene is the first example of
truly two-dimensional crystals, in contrast with numerous {\it
quasi}-two-dimensional crystals known before. This opens many
interesting questions concerning thermodynamics, lattice dynamics
and structural properties of such systems which are, however,
beyond the scope of this paper. Furthermore, single-layer graphene
provides first experimental realization of a two-dimensional
massless Dirac fermion system. The analogy with the quantum field
theory proved crucial for understanding of graphene unusual
electronic properties, such as anomalous QHE, absence of the
Anderson localization, inefficiency of scattering by point
defects, etc. The bilayer graphene has a very unusual gapless
parabolic spectrum providing an example of the system with
electron wave equation different from both Dirac and
Schr\"{o}dinger ones.

\textit{Acknowledgements}. We are thankful to Andre Geim, Maria
Vozmediano, and Leonid Levitov for helpful discussions. This work
was supported by the Stichting voor Fundamenteel Onderzoek der
Materie (FOM), the Netherlands and by the Royal Society, UK.

\end{document}